\documentclass[runningheads]{llncs}
\usepackage[T1]{fontenc}
\usepackage{graphicx}
\usepackage{multirow}
\usepackage{booktabs}

\usepackage{listings}
\usepackage{xcolor}
\usepackage{tcolorbox}

\definecolor{darkgray}{gray}{0.3}

\begin{document}
\title{Evaluating Large Language Models for Decision-Making in Agent-Based Urban Mobility Simulations}

\titlerunning{Evaluating LLMs for Decision-Making in Agent-Based Mobility Simulations}

\author{
    Bruno Cascaes Alves\inst{1}\and
    Míriam Blank Born\inst{1}\and \\
    Ulisses Gilioli Francescatto Júnior\inst{2}\and
    Felipe Moura Goulart\inst{3}\and \\
    Letícia Brandão Caldas\inst{3}\and
    Marilton Sanchotene de Aguiar\inst{1}
}

\authorrunning{B. C. Alves et al.}

\institute{
Postgraduate Program in Computing \and
Undergraduate in Computer Science \and
Undergraduate in Computer Engineering\\
Federal University of Pelotas, Pelotas, Brazil\\
\email{\{bcalves,mbborn,ugfjunior,fmgoulart,lbcaldas,marilton\}@inf.ufpel.edu.br}
}

\maketitle              
\begin{abstract}
Urban mobility modeling faces challenges in representing decision-making in dynamic environments. Although Multi-Agent Systems are widely used, rule-based approaches rely on fixed heuristics that limit adaptive behavior. This work investigates the integration of Large Language Models (LLMs) as decision-making components in multi-agent simulations. We propose a hybrid architecture that connects the GAMA platform to an external LLM-based module through an API, enabling agents to determine whether route replanning is necessary. Rather than replacing routing algorithms, the LLM serves as a decision layer that guides replanning behavior. The approach incorporates persistent memory, allowing past interactions to influence future decisions and promote behavioral consistency. We compare rule-based and LLM-assisted approaches across multiple road-blockage scenarios and population scales. Results indicate that LLM-enabled agents exhibit greater adaptability and contextual awareness, particularly in scenarios with higher route flexibility. Memory influences performance and behavioral consistency, with effects varying across configurations. Overall, LLMs serve as complementary cognitive layers that enrich behavioral representations in urban mobility simulations and hold potential for modeling complex decision-making in spatial multi-agent systems.

\end{abstract}

\section{Introduction}

Urban mobility management faces the challenge of predicting and mitigating the impacts of events in increasingly saturated transportation networks \cite{mavlutova2023urban}. The efficacy of interventions depends on the understanding that traffic dynamics emerge from individual behavioral decisions made by agents in dynamic environments under uncertainty \cite{bazzan2013traffic}. In this context, Artificial Intelligence (AI) emerges as a component for increasing realism of experiments, enabling the incorporation of adaptation and contextual reasoning in urban mobility analysis \cite{russell2020artificial}.

Multi-Agent Systems (MAS) simulations have consolidated as a predominant approach for representing individuals and analyzing their impacts on urban systems. This approach enables the investigation of mobility dynamics through the interaction of autonomous agents in dynamic environments, capable of making decisions to achieve individual or collective goals~\cite{wooldridge2009multiagent}. However, traditional modeling often relies on fixed heuristics and predefined rules, limiting behavioral adaptability in high-contextual-variability scenarios.

To mitigate these limitations, Large Language Models (LLMs) emerge as a promising alternative for enriching decision-making in agent-based systems. Models such as GPT~\cite{achiam2024gpt} and Gemini~\cite{anil2025gemini} exhibit reasoning and semantic capabilities that enable the processing of contextual information \cite{ozdemir2023quick}. By integrating these models into MAS, agents can exhibit greater behavioral flexibility and make more context-sensitive decisions aligned with environmental conditions.

Recent studies demonstrate advances in LLM-based Agentic frameworks, in which these models serve as system cores capable of planning and iterating on their own actions. These frameworks, including Agno \cite{sachenkova2026agentic} and CrewAI \cite{venkadesh2024unlocking}, utilize memory mechanisms and integration with external tools, promoting greater consistency and continuity in decision-making over time. However, researchers primarily evaluate these approaches in textual domains and rarely explore them in multi-agent simulations that heavily rely on geospatial constraints.

The GAMA platform \cite{taillandier2019gama} stands out as one of the most robust agent-based simulation environments, with native support for geographic data. Despite its efficiency in spatial modeling, the platform lacks native mechanisms for integrating language models into agents' decision-making processes, limiting the representation of more adaptive, context-aware behaviors.

Given this scenario, this work proposes integrating an LLM-based decision module into a GAMA-based urban mobility simulation environment. Unlike purely rule-based approaches, the proposal introduces a context-driven mechanism triggered by events, in which the LLM determines whether to maintain the agents' trajectory or trigger route recalculation. The objective is to explore how this decision layer influences agent behavior under dynamic conditions of urban mobility. The central hypothesis is that this mechanism enhances behavioral adaptability while preserving simulation consistency.


\section{Theoretical Background}

This section presents the fundamental concepts for this work, including Multi-Agent Systems, Large Language Models, Agentic AI, and the GAMA Platform.

Multi-Agent Systems consist of multiple autonomous entities that interact within a shared environment. Each agent operates based on local information, and global system behavior emerges from their interactions. This decentralized structure is particularly suitable for modeling complex systems where collective dynamics arise from individual decision-making processes \cite{dorri2018mas_survey}.

Large Language Models are neural networks trained on large-scale text corpora that can perform language understanding and generation tasks. Beyond their traditional use in language understanding and generation tasks, LLMs can also support decision-making processes by providing context-aware interpretations of structured inputs. When integrated into decision pipelines, LLMs can act as reasoning components that transform contextual information into structured outputs, enabling their use in systems that require adaptive responses \cite{zhao2023survey}.

Agentic AI comprises autonomous systems that perceive environmental states and execute actions to achieve specific objectives. Instead of relying solely on predefined rules, Agentic AI incorporates context-aware decision mechanisms that integrate memory and external model feedback, which directly influence their actions during simulation runtime \cite{sapkota2026agentic_ai}. In this work, we model agents as components of a multi-agent simulation system and augment their decision-making process with LLMs, enabling more adaptive behavior in dynamic environments.

Finally, GAMA is an open-source platform for agent-based modeling and simulation with strong support for spatially explicit systems. It enables the representation of heterogeneous agents operating in environments based on geographic information systems (GIS), supporting large-scale simulations of complex systems. GAMA provides the GAML modeling language and tools for defining agent behaviors, spatial interactions, and environment dynamics \cite{taillandier2019gama}. 

\section{Related Work}

This section presents the main research areas relevant to this work, organized into two perspectives: agent-based urban mobility simulation in spatial environments and the use of LLM-based Agentic frameworks as decision-making components in intelligent systems.

The literature reports several applications of multi-agent systems for urban mobility in geospatial environments. For instance, \cite{olszewski2019spatiotemporal} uses GAMA to investigate social dynamics in smart cities, modeling daily routines and interactions between urban infrastructures. In the context of transportation planning, the authors of \cite{grignard2018baseline} consider the GAMA platform to model mobility patterns and evaluate the integration of new transport modes. Additionally, studies such as \cite{barbosa2024escape} explore evacuation scenarios with heterogeneous agents. Despite these contributions, such approaches primarily rely on rule-based models, limiting their ability to represent adaptive behaviors under dynamic conditions. 

Advances in LLMs have enabled new forms of decision support in intelligent systems. Frameworks such as CrewAI \cite{venkadesh2024unlocking} and AutoGen \cite{wu2023autogen} provide mechanisms to structure LLM-based workflows, enabling coordination among components, integration with external tools, and the use of persistent memory. These approaches have demonstrated promising results across domains such as software engineering and collaborative task automation. However, these approaches are primarily designed for language-centered environments and rarely consider spatial constraints in MAS simulations.

Based on the related work, we identify a lack of studies that explore LLM-based decision mechanisms in spatially explicit multi-agent simulations where agents must respond to real-time changes in urban mobility while respecting road network constraints. To address this gap, we incorporate an LLM-based decision layer into a GAMA urban mobility simulation. The proposed component determines whether agents should maintain their current route or trigger recalculation based on contextual information and memory, enabling the investigation of context-sensitive decision-making in dynamic mobility scenarios.

\section{Methodology}

\begin{figure}[ht!]
    \centering \includegraphics[width=.8\textwidth]{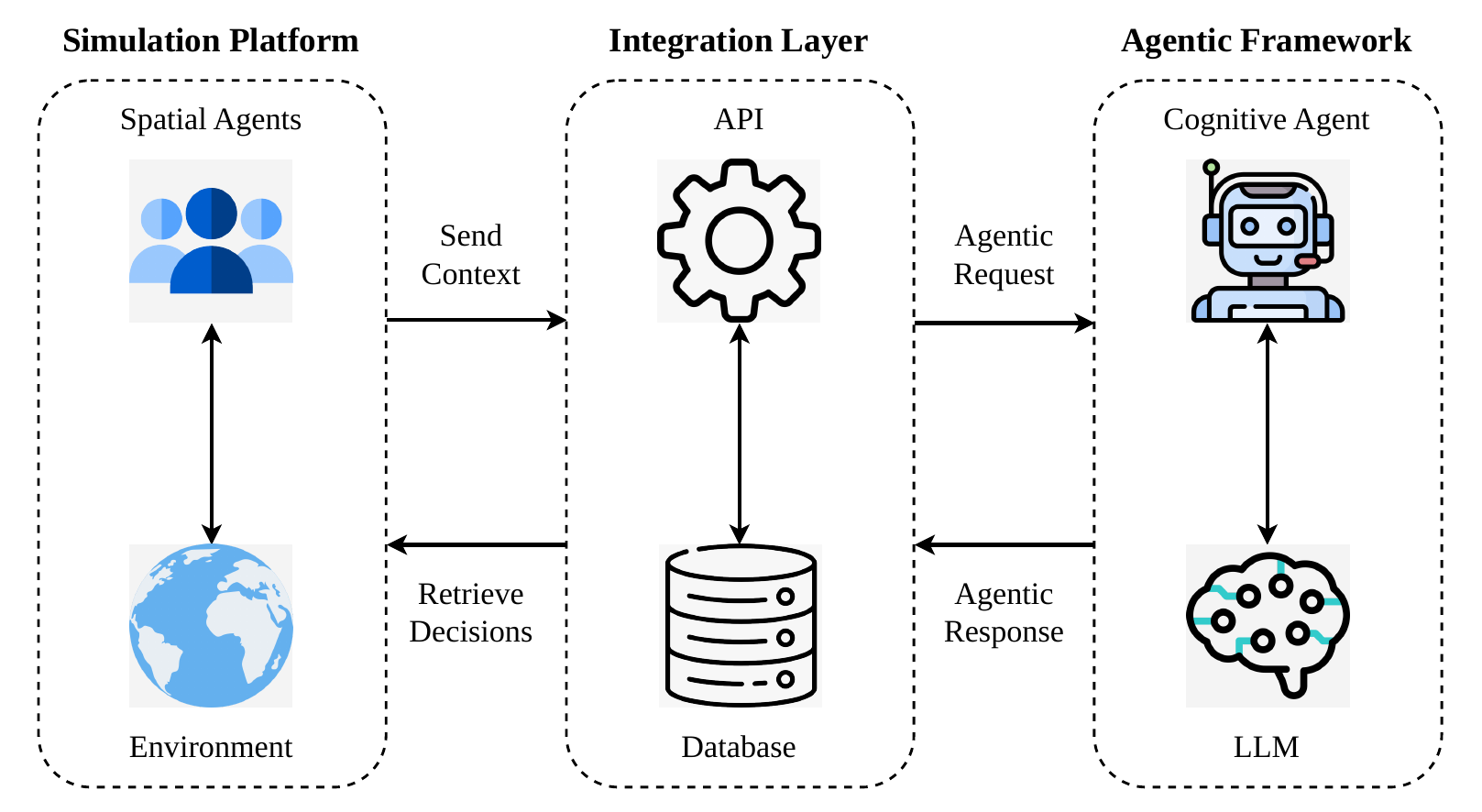} 
    \caption{Overview of the proposed architecture integrating an LLM-based decision layer into the simulation platform.}
    \label{fig:architecture}
\end{figure}

This section describes the experimental framework used to evaluate LLM-based agents in an urban mobility simulation. Building on previous work on the application of LLM-based agents in an urban mobility scenario \cite{alves2026integrating}, which investigated transportation mode choice among heterogeneous agent profiles, the present study shifts the focus to in-transit decision-making. Specifically, it examines how different language models, memory configurations, and environmental conditions influence route-adaptation decisions in response to dynamic events in the road network. The evaluation focuses on mobility-related behaviors such as route selection and adaptation to disruptions encountered throughout the simulation. Figure~\ref{fig:architecture} presents an overview of the architecture, organized into three main components:

\begin{description}
    \item[Simulation Platform:] Represents the GAMA-based environment and its agents. The environment includes buildings and a road network. Agents travel between origins and destinations using shortest-path algorithms on weighted graphs, where traffic conditions affect travel costs. During the simulation, they monitor congestion and road events and receive alerts that may influence LLM-based decisions.

    \item[Integration Layer:] Acts as an intermediary between the simulation platform and the Agentic framework. It handles API communication, retrieves contextual information from GAMA, and forwards it to the LLM-based Agentic module. It also manages a relational database that stores simulation and round states, agent attributes, and decision logs, ensuring traceability and supporting memory mechanisms.

    \item[Agentic Framework:] Implements the cognitive layer through the Agno framework, which orchestrates contextual reasoning, memory management, and LLM-based decision-making. The module processes contextual information and returns a structured decision on whether an agent should maintain its current route or trigger a recalculation, updating the agent's behavior within the GAMA platform.
 \end{description}

The following subsections describe the components that support the experimental evaluation: i) Environment and Agent Modeling; ii) Integration and Persistence Layer; and, finally, iii) LLM-based Agentic Module.

\subsection{Environment and Agent Modeling}

\begin{figure}[ht!]
    \centering \includegraphics[width=.9\textwidth]{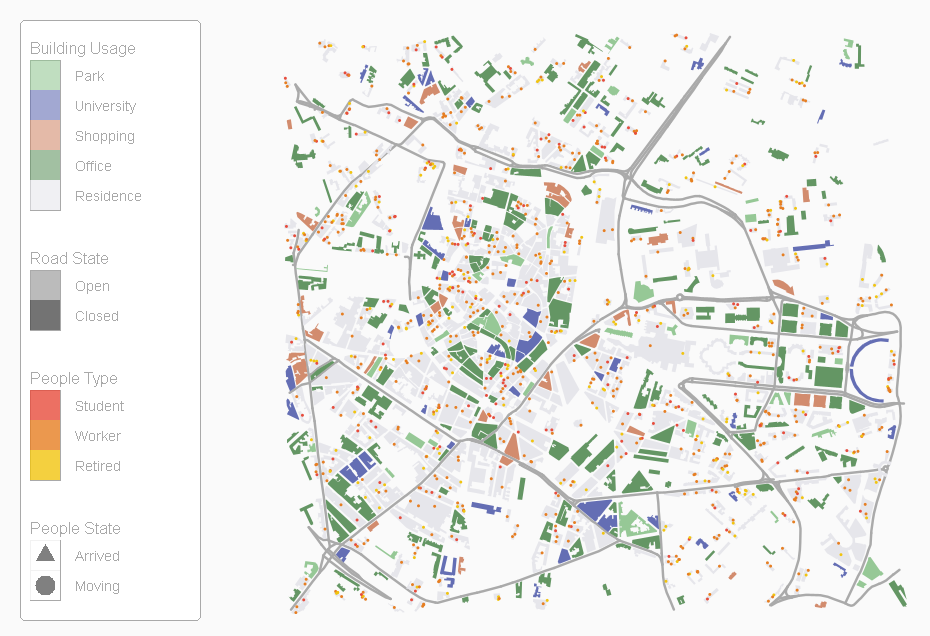} 
    \caption{Spatial representation at the beginning of a simulation in GAMA.}
    \label{fig:env-initial-state}
\end{figure}

This study extends the urban mobility model based on the \textit{Traffic and Pollution} example from the GAMA platform. In the original model, mobility dynamics depend on vehicle density, and agents move using shortest-path algorithms over a weighted road network derived from geographic data. Although the original model includes pollution-related mechanisms, this work does not incorporate them and focuses instead on the agents' mobility decision-making process. Figure~\ref{fig:env-initial-state} illustrates the spatial representation of the simulation in the GAMA platform, highlighting agent interaction with urban infrastructure. 

The model represents agents as colored circles according to their profile (student, worker, or retired) and as triangles once they reach their destinations. Each agent is also associated with attributes such as digital trust level and accumulated travel time, which influence its response to traffic events during the simulation.

For comparison purposes, we implement a rule-based baseline model that does not incorporate language models. When agents approach a blocked road segment, they receive an alert and decide whether to continue on their route or recalculate it. This decision relies on a threshold-based heuristic score $S$, composed of additive contextual factors: $S = F_{dist} + F_{cong} + F_{time} + \Delta_{trust}$, where $F_{dist}$, $F_{cong}$, and $F_{time}$ represent the influence of event proximity, local congestion, and accumulated travel time, respectively. The term $\Delta_{trust}$ models behavioral differences between agent profiles through a linear transformation centered on digital trust level, modulating sensitivity to route changes. The model triggers route recalculation when $S > 0.5$.

To overcome the limitations of the rule-based model, we replace the heuristic mechanism with an LLM-based decision module integrated via API. In this approach, when an agent receives an alert, it sends its current state to the cognitive module, which decides whether to trigger route recalculation. The LLM does not perform route computation; instead, it serves solely as a decision layer that determines when to run the traditional routing algorithm.

\subsection{Integration and Persistence Layer}

To integrate the simulation platform with the agentic framework, we develop a layer implemented as a Python API using FastAPI~\cite{ramirez2018fastapi}. This layer manages communication between system components through asynchronous HTTP requests, allowing multiple agents to request decisions concurrently without blocking simulation execution.

The system includes a relational database to store simulation information, including agent states, environmental conditions, and generated decisions. Each decision record contains the agent's context at the time of the event, including occupation, digital trust level, travel time, and location, along with the LLM's response, enabling detailed analysis and traceability. Beyond experiment logging, the persistence layer supports memory retrieval by storing previous agent experiences and decision outcomes. Retrieved records can be incorporated into the contextual information sent to the LLM, allowing decisions to consider both current environmental conditions and relevant past interactions.

The integration follows an event-driven model. For instance, when a road interruption alert occurs, the agent sends its contextual state to the API, which forwards the request to the LLM-based agentic module. The module evaluates the context and returns a structured decision indicating whether to trigger route recalculation. This architecture decouples cognitive processing from spatial simulation, ensuring experimental flexibility while preserving the simulation platform's execution model.

\subsection{LLM-based Agentic Module}

To enable adaptive decision-making, we implement an LLM-based module using the Agno framework. Unlike rule-based approaches, the module analyzes the agent's and environmental contexts to determine whether route recalculation is necessary. We chose Agno for its flexibility and support for semantic memory. For each decision, the Agno system stores the observed context, selected action, and model-generated reasoning. When memory is enabled, semantically similar experiences can be retrieved and incorporated into the reasoning process, allowing previous outcomes to influence subsequent actions.

The decision process is triggered whenever an agent receives a traffic alert. Agno generates a prompt from agent attributes (e.g., occupation and digital trust level), current travel conditions (e.g., fatigue level), and environmental factors (e.g., congestion level). Listing~\ref{lst:1} shows a simplified version of the prompt; the actual implementation includes additional instructions to enforce structured outputs. The LLM returns a JSON response indicating whether the agent should maintain its route or trigger recalculation, along with a brief justification. Invalid responses are validated and, when necessary, replaced through a fallback mechanism. This approach enables the analysis of context and memory-aware decisions while preserving the original routing algorithms.

Furthermore, it is possible to incorporate additional contextual variables, external information sources, or auxiliary tools into the system during reasoning. This flexibility makes the proposed approach suitable for exploring increasingly complex decision-making processes in urban mobility simulations.

\renewcommand{\thelstlisting}{\arabic{lstlisting}}
\begin{lstlisting}[caption={Simplified prompt illustrating the contextual information provided to LLM-based agents for route-replanning decisions.}, captionpos=b, label={lst:1}]
You are a driver traveling through a city. A digital traffic 
alert indicates a possible road blockage ahead. Based on the 
driver's profile and the current travel conditions, decide 
whether to maintain the current route or recalculate it.

Profile:                        Context:
- Occupation: Worker            - Weather: Rainy
- Fatigue Level: Medium         - Traffic Congestion: High
- Trust Level: High             - Distance to blockage: Near

Return ONLY a JSON object in the following format:

{
    "action": "RECALCULATE | STAY",
    "reasoning": "<brief_explanation>"
}

\end{lstlisting}

\section{Results and Discussion}

This section presents the experimental evaluation of the proposed architecture, focusing on how LLMs influence agent behavior in urban mobility simulations. Rather than a direct performance comparison with a rule-based model, the analysis aims to understand how context-aware decision mechanisms, particularly the use of memory, affect agent adaptation in response to dynamic events.

\subsection{Experimental Setup}

To assess the impact of incorporating LLMs into the decision-making process, we conducted experiments comparing three approaches: i) a rule-based baseline model, ii) an LLM-assisted model using Google's Gemini 2.5 Flash-Lite, and iii) an LLM-assisted model using OpenAI's GPT-4o Mini. For the LLM-based models, we evaluated configurations with and without persistent memory. We selected GPT and Gemini models for their favorable balance of performance, cost, and inference speed in large-scale agent-based simulations. Also, we run all experiments with a temperature setting of 1.0, balancing response consistency and behavioral variability in the agents' decisions.

We conducted the experiments using the GAMA platform, considering a road network derived from geographic data and heterogeneous agents classified into three profiles: students (20\%), workers (60\%), and retirees (20\%). Each agent has attributes such as a digital trust level and a dynamic travel state that influence its decision-making. Each simulation run consists of 540 cycles, each representing 10 seconds in the simulated environment, for a total of approximately 90 minutes. During the simulation, when a road interruption alert occurs, each agent decides whether to maintain its current route or trigger route recalculation using traditional shortest-path algorithms. In the LLM-assisted models, an external cognitive module handles this decision.

The system introduces interruption events dynamically, starting at cycle 120 and being removed at cycle 420, allowing the observation of both adaptation and recovery phases of the system. We implemented two scenarios: Localized, characterized by a point-like interruption with greater availability of alternative routes, and Extended, characterized by the disruption of significant road segments, substantially reducing rerouting options. For each scenario, we consider two population scales (500 and 1000 agents) to evaluate the impact of density on collective behavior.

We run each experimental configuration with 10 independent random seeds and report mean values and, when applicable, standard deviations. We consider the following metrics: i) Replanning Rate, the percentage of route recalculation decisions; ii) Arrival Rate, the percentage of agents that reach their destinations; iii) Stuck Time Ratio, the proportion of total travel time spent in congestion; iv) Simulation Time, the total simulation execution time in seconds; and v) Error Rate, the percentage of responses that could not be processed by the system, including invalid JSON outputs. Together, these metrics enable the analysis of both individual performance and emergent collective behavior in the network.

\subsection{Localized Blockage Scenario}

\begin{figure}[ht!]
    \centering \includegraphics[width=.9\textwidth]{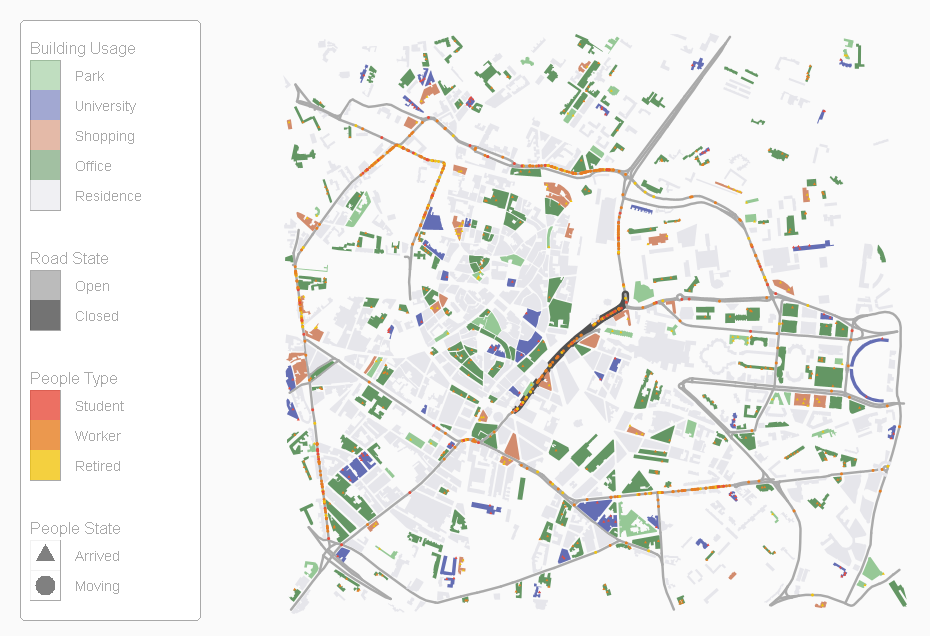} 
    \caption{Spatial representation during a localized blockage event in GAMA.}    \label{fig:env-local}
\end{figure}

Results for the localized blockage scenario show that incorporating LLMs significantly changes agent decision dynamics, particularly in local adaptation and spatial redistribution across the network. As shown in Figure~\ref{fig:env-local}, LLM-assisted agents tend to avoid critical regions earlier, leading to a more balanced distribution across the road network. This behavior arises not only from more frequent replanning but also from the ability to interpret contextual information and adjust decisions accordingly.

Table~\ref{tab:local-blockage} shows that LLM-assisted approaches generally achieve higher arrival rates and comparable or lower congestion levels when compared to the rule-based baseline. For 500 agents, the arrival rate increases from 70\% to approximately 90\% (OpenAI with memory). At the same time, the Stuck Time Ratio varies across configurations, decreasing to 11.06\% for Gemini with memory but increasing to 14.17\% for OpenAI with memory. For 1000 agents, the arrival rate increases from 53\% to 62\% (OpenAI with memory), and the Stuck Time Ratio decreases to around 5\% (Gemini with memory).

Memory affects the models differently. For instance, with 500 agents, OpenAI found that memory reduces replanning (from 94.1\% to 82.8\%) while increasing arrival rates (from 83.5\% to 89.7\%), indicating changes in decision consistency over time. However, it also increases the Stuck Time Ratio in some configurations (from 9.4\% to 14.2\%), suggesting potential trade-offs between successful arrival and congestion. For Gemini, memory reduces the Stuck Time Ratio (from 13.2\% to 11.1\%) and increases arrival rates, indicating more effective utilization of prior experience in congestion avoidance. The observed changes in replanning rates suggest that memory serves as a behavioral stabilizer, reducing excessive reactive decisions and favoring responses more consistent with prior experiences.

\begin{table}[ht!]
    \centering
    \caption{Comparative analysis of semantic memory impact on affected agents under a local blockage scenario.}
    \label{tab:local-blockage}
    \setlength{\tabcolsep}{4pt}
    \renewcommand{\arraystretch}{1.2}
    \resizebox{.9\textwidth}{!}{%
    \begin{tabular}{cccccc}
    \hline
    \textbf{Agents} & \textbf{Model} & \textbf{Memory} & \textbf{Replanning (\%)} & \textbf{Arrival (\%)} & \textbf{Stuck Ratio (\%)} \\ \hline
    \multirow{5}{*}{500}	& Rule-Based & --- & 63.06 & 70.00 & 15.03 \\
     & Gemini-Assisted & No & 66.58 & 78.25 & 13.24 \\
     & Gemini-Assisted & Yes & 63.33 & 79.44 & 11.06 \\ 
     & OpenAI-Assisted & No & 94.10 & 83.45 & \textbf{9.43} \\ 
     & OpenAI-Assisted & Yes & 82.76 & \textbf{89.66} & 14.17 \\ \hline
    
    \multirow{5}{*}{1000} & Rule-Based & --- & 68.37 & 53.00 & 10.74 \\ 
     & Gemini-Assisted & No & 80.39 & 54.81 & 7.55 \\ 
     & Gemini-Assisted & Yes & 85.92 & 61.50 & \textbf{4.94} \\ 
     & OpenAI-Assisted & No & 97.42 & 58.88 & 6.41 \\
     & OpenAI-Assisted & Yes & 92.91 & \textbf{62.06} & 5.06 \\ \hline
    \end{tabular}
    }
\end{table}

\begin{figure}[ht!]
    \centering \includegraphics[width=.75\textwidth]{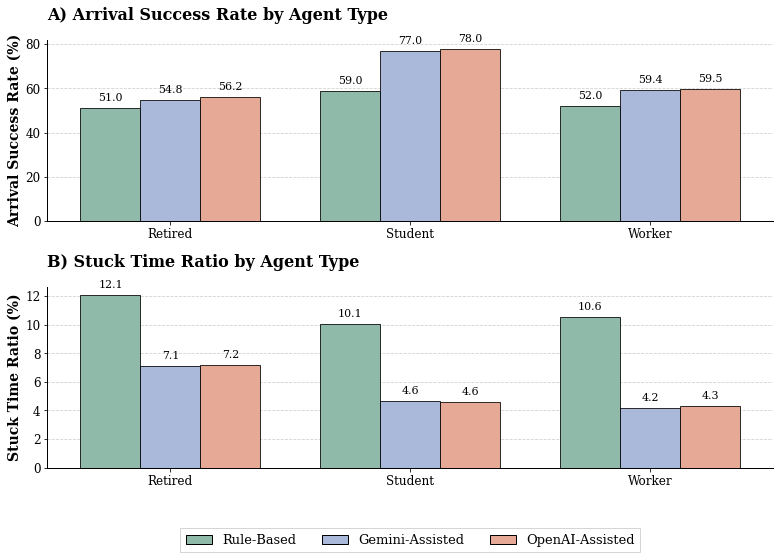} 
    \caption{Arrival rate and stuck time ratio by agent profile for 1000 agents under the localized blockage scenario.}
    \label{fig:results-localized}
\end{figure}

The per-profile analysis for 1000 agents, shown in Figure~\ref{fig:results-localized}, further highlights behavioral differences across agent types. Students exhibit higher arrival rates with LLM-assisted approaches, whereas retirees show smaller variation. These results suggest that LLM-based decision-making induces heterogeneous responses across agent profiles, contributing to more diverse system-level dynamics.

Overall, the results indicate that LLM-assisted decision-making does not act as a strict performance optimizer compared to rule-based approaches, but instead generates heterogeneous behavioral dynamics under local blockage conditions. Model choice and memory jointly shape how agents adapt to congestion, with memory serving to incorporate prior experience into decision-making and modulate agent responses over time.

\subsection{Extended Blockage Scenario}

\begin{figure}[ht!]
    \centering \includegraphics[width=.9\textwidth]{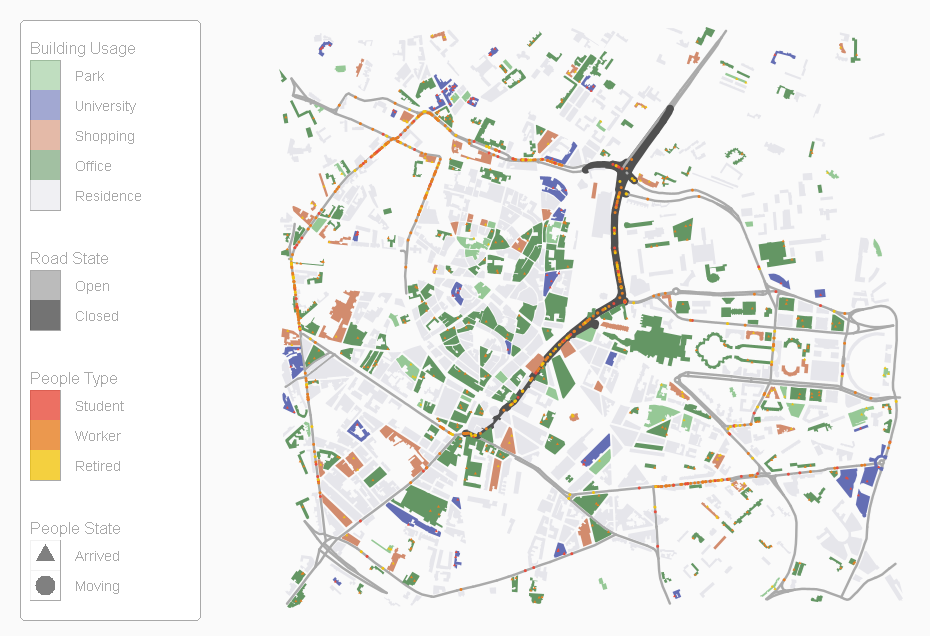} 
    \caption{Spatial representation during an extended blockage event in GAMA.}
    \label{fig:env-extended}
\end{figure}

In the extended blockage scenario, the results highlight the limitations of individual decision-based adaptation when the network structure imposes strong constraints. As shown in Figure~\ref{fig:env-extended}, disruptions to critical roads substantially reduce route alternatives, concentrating traffic on a few links. In this context, LLM-based decision-making still affects local agent behavior but has a reduced impact on overall system performance.

Table~\ref{tab:extended-blockage} shows that LLM-assisted approaches consistently reduce the stuck time ratio compared to the rule-based baseline, although gains in arrival rate remain limited. For 500 agents, the Stuck Time Ratio drops from 8.01\% to 1.86\% (OpenAI with memory), with similar reductions for 1000 agents, where values fall to around 2\% across LLM-assisted configurations. Despite these improvements, arrival rates increase only modestly for 500 agents (up to 72.13\% with Gemini and memory) and remain close to the baseline for 1000 agents (around 40\%). These results suggest that, under extended blockage conditions, LLMs improve local congestion avoidance but are constrained by global network limitations in translating these gains into overall performance.

\begin{table}[ht!]
    \centering
    \caption{Comparative analysis of semantic memory impact on affected agents under an extended blockage scenario.}
    \label{tab:extended-blockage}
    \setlength{\tabcolsep}{4pt}
    \renewcommand{\arraystretch}{1.2}
    \resizebox{.9\textwidth}{!}{%
    \begin{tabular}{cccccc}
    \hline
    \textbf{Agents} & \textbf{Model} & \textbf{Memory} & \textbf{Replanning (\%)} & \textbf{Arrival (\%)} & \textbf{Stuck Ratio (\%)} \\ \hline

    \multirow{5}{*}{500} & Rule-Based & --- & 70.62 & 67.00 & 8.01 \\
     & Gemini-Assisted & No & 83.60 & 63.72 & 3.89 \\
     & Gemini-Assisted & Yes & 80.83 & \textbf{72.13} & 4.19 \\
     & OpenAI-Assisted & No & 97.20 & 62.74 & 2.61 \\
     & OpenAI-Assisted & Yes & 95.58 & 62.69 & \textbf{1.86} \\ \hline
    
    \multirow{5}{*}{1000} & Rule-Based & --- & 82.07 & 40.00 & 6.14 \\ 
     & Gemini-Assisted & No & 88.73 & \textbf{41.90} & 3.14 \\ 
     & Gemini-Assisted & Yes & 90.94 & 41.41 & 2.53 \\
     & OpenAI-Assisted & No & 98.62 & 39.53 & 2.22 \\
     & OpenAI-Assisted & Yes & 96.64 & 40.69 & \textbf{2.15} \\ \hline
    \end{tabular}
    }
\end{table}

Memory exhibits model-dependent effects. For OpenAI-assisted agents, it reduces the time spent stuck. Still, it has a negligible impact on arrival rates, indicating a role focused on local behavioral refinement rather than global performance improvement. For Gemini-assisted agents, memory contributes more noticeably to improvements in arrival rate in the 500-agent scenario (from 63.72\% to 72.13\%), while also slightly increasing replanning. However, for 1000 agents, memory effects are marginal in both models, yielding only small reductions in congestion without significantly altering arrival rates. Overall, these results suggest that memory serves as a stabilizing mechanism whose impact diminishes as environmental constraints intensify.

\begin{figure}[ht!]
    \centering \includegraphics[width=.8\textwidth]{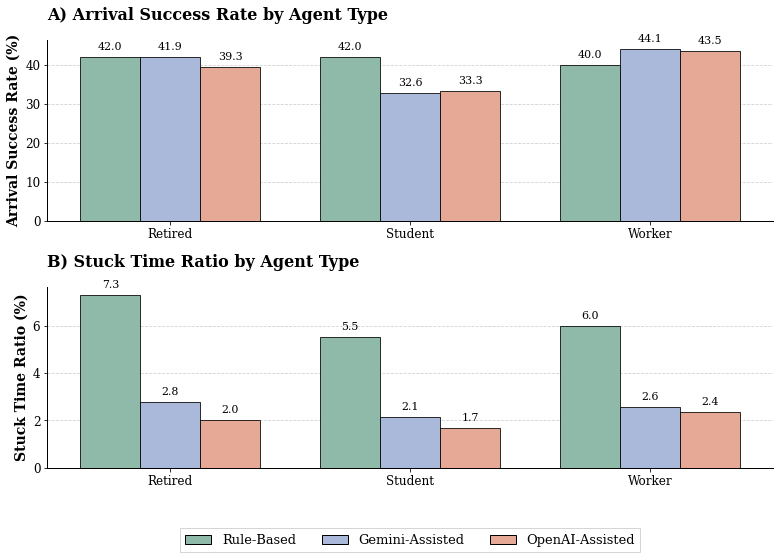} 
    \caption{Arrival rate and stuck time ratio by agent profile for 1000 agents under the extended blockage scenario.}
    \label{fig:results-extended}
\end{figure}

The per-profile analysis for 1000 agents, shown in Figure~\ref{fig:results-extended}, further indicates a reduction in behavioral differences across agent types when compared to the localized scenario. Unlike the previous case, where individual profiles exhibited more pronounced variation under LLM-assisted decision-making, here we observe stronger convergence in both arrival rates and congestion-related metrics among retirees, students, and workers. These results suggest that, under extended blockage conditions, system dynamics are increasingly dominated by structural network constraints, limiting the influence of individual agent characteristics and reducing the impact of LLM-mediated behavioral heterogeneity.

\subsection{Computational Cost and Robustness Analysis}

Table~\ref{tab:cost-robustness} summarizes the computational cost and robustness of the evaluated approaches in terms of simulation time and error rate. Overall, the results indicate that incorporating LLM-based decision-making results in a substantial increase in computational cost compared to the rule-based baseline. This increase is consistent across both population scales, reflecting the additional overhead introduced by external model inference and communication between components. In contrast, the rule-based model remains significantly more efficient due to its purely local and deterministic execution.

Despite this overhead, error rates remain low across all LLM-based configurations, indicating that robustness is not significantly affected by LLM integration. Even in more complex configurations, failures are rare, suggesting stable interaction between the simulation environment and the external decision module. These results highlight a trade-off: while LLM integration increases computational cost, it preserves robustness and enables richer decision-making within the simulation framework.

\begin{table}[ht!]
    \centering
    \caption{Computational cost and robustness analysis of the evaluated approaches, reporting average simulation time and error rate across simulations.}
    \label{tab:cost-robustness}
    \setlength{\tabcolsep}{10pt}
    \renewcommand{\arraystretch}{1.2}
    \resizebox{.9\textwidth}{!}{%
    \begin{tabular}{ccccc}
    \hline
    \textbf{Agents} & \textbf{Model} & \textbf{Memory} & \textbf{Simulation Time (s)} & \textbf{Error Rate (\%)} \\ \hline
    \multirow{5}{*}{500} & Rule-Based & --- & $21.92 \pm 5.78$ & --- \\
     & Gemini-Assisted & No & $79.40 \pm 16 .04$ & 0.58 \\
     & Gemini-Assisted & Yes & $98.50 \pm 10.61$ & 1.17 \\
     & OpenAI-Assisted & No & $102.30 \pm 24.27$ & 0.00 \\
     & OpenAI-Assisted & Yes & $111.50 \pm 12.62$ & 0.00 \\ \hline
    
    \multirow{5}{*}{1000} & Rule-Based & --- & $58.40 \pm 7.79$ & --- \\
     & Gemini-Assisted & No & $152.50 \pm 24.62$ & 0.49 \\
     & Gemini-Assisted & Yes & $219.25 \pm 34.28$ & 1.25 \\
     & OpenAI-Assisted & No & $184.80 \pm 22.60$ & 0.00 \\
     & OpenAI-Assisted & Yes & $239.67 \pm 28.50$ & 0.00  \\ \hline
    \end{tabular}
    }
\end{table}

\section{Conclusion and Future Work}

This study evaluates an architecture that integrates an LLM-based Agentic framework as a decision module within multi-agent urban mobility simulations. We propose a hybrid setup connecting the GAMA platform to the Agno framework, replacing rule-based heuristics with context-sensitive reasoning enhanced by semantic memory while preserving spatial consistency.

The results suggest that LLM-assisted decision-making changes agent adaptation patterns. In localized blockages, agents improve traffic redistribution and arrival rates. In large-scale disruptions, road constraints limit global gains, though local congestion avoidance improves. Memory acts as a stabilizer, reducing overly reactive behavior. Also, LLMs operate more effectively as complementary cognitive layers than as replacements for routing algorithms. As decision filters, they improve the interpretation of context and responses to unexpected events through semantic reasoning, resulting in more realistic behaviors. However, this raises computational overhead, as shown in execution time results, highlighting the need for optimization in large-scale scenarios.

Overall, the approach demonstrates the potential of LLM-based agent architectures to enhance multi-agent systems and enable more expressive urban mobility modeling. Future work will address scalability, refine memory mechanisms for greater efficiency, and extend the evaluation to additional decision-making tasks beyond route recalculation. The findings support hybrid AI architectures as a promising direction for data-driven intelligence in complex systems.

\section*{Acknowledgements}

This study was financed in part by the Coordenação de Aperfeiçoamento de Pessoal de Nível Superior -- Brasil (CAPES) -- Finance Code 001. This work was supported by the Kunumi Institute. The authors thank the institution for its financial support and commitment to advancing scientific research.

Generative Artificial Intelligence tools, including Gemini and Grammarly, were employed to refine the linguistic quality during the preparation of this manuscript. The authors did not fabricate data, manipulate experimental results, generate fictitious references, or misrepresent the scientific interpretation. All content was reviewed and verified by the authors, who assume full responsibility for the final version of the manuscript.

\bibliographystyle{splncs04}
\bibliography{bracis.bib}

\end{document}